\documentclass[a4paper,11pt]{article}
\usepackage{pos}
\pdfoutput=1   
\usepackage{mathrsfs,latexsym}
\usepackage{graphicx}
\usepackage{xcolor}
\usepackage{appendix}
\usepackage{fontawesome} 
\usepackage{bm}
\usepackage{times}
\usepackage{float}
\usepackage{wrapfig}
\usepackage[utf8]{inputenc} 
\usepackage[T1]{fontenc} 
\usepackage{hyperref}
\newcommand{\eq}[1]{Eq.~(\ref{#1})}

\newcommand{\fig}[1]{figure~\ref{#1}}

\newcommand{\tr}{\mbox{Tr}}

\newcommand{\ben}{\begin{enumerate}}
\newcommand{\een}{\end{enumerate}}
\newcommand{\bit}{\begin{itemize}}
\newcommand{\eit}{\end{itemize}}
\newcommand{\beq}{\begin{equation}}
\newcommand{\eeq}{\end{equation}}
\newcommand{\bsa}{\begin{subequations}\begin{eqnarray}}
\newcommand{\esa}{\end{eqnarray}\end{subequations}}
\newcommand{\bea}{\begin{eqnarray}}
\newcommand{\eea}{\end{eqnarray}}
\newcommand{\bean}{\begin{eqnarray*}}
\newcommand{\ean}{\end{eqnarray*}}
\newcommand{\nn}{\nonumber \\}
\newcommand{\non}{\nonumber}

\title{Hybrid static potentials from Laplacian Eigenmodes}

\author*[a]{Roman~H\"ollwieser}
\author[a]{Francesco Knechtli}
\author[a]{Tomasz Korzec}
\author[b]{Michael Peardon}
\author[a]{Juan Andrés Urrea-Niño}
\affiliation[a]{Department of Physics, University of Wuppertal, Gau{\ss}strasse 20, 42119 Germany}
\affiliation[b]{School of Mathematics, Trinity College Dublin, Ireland}

\emailAdd{hoellwieser@uni-wuppertal.de}

\abstract{We present a method for computing hybrid static quark-antiquark potentials in lattice QCD based on Laplace trial states. They are formed by eigenvector components of the covariant lattice Laplace operator and their covariant derivatives. The new method does not need complicated gauge link paths between the static quarks and makes off-axis separations easily accessible. We show first results for $\Sigma$ and $\Pi$ together with their excited states on a dynamical ensemble.}

\FullConference{The 40th International Symposium on Lattice Field Theory (Lattice 2023)\\
July 31st - August 4th, 2023\\
Fermi National Accelerator Laboratory\\}

\begin{document}
\maketitle

\section{Introduction}

In~\cite{Hollwieser:2022doy} we have introduced a new operator, namely "Laplace trial states", which replace the spatial Wilson line in a classical Wilson loop with a weighted sum of eigenvector pairs of the 3D lattice Laplace operator. In the case of the static potential we get an improvement for the static energies, which reach their plateau values at earlier temporal distances and we basically get off-axis distances for free. Here, we want to apply this technique to compute static-hybrid potentials, where the gluonic excitations are realized via covariant derivatives of individual eigenvectors. The technology developed in this article for hybrid potentials can be applied to multi-quark potentials which provide insight into the internal structure of exotic configurations of static sources with non-trivial spin and isospin \cite{Bicudo:2022cqi}, as well as static-light potentials.

\section{Methods}\label{sec:meth}

Static potentials are classically measured via Wilson loops, which arise from correlations in time of trial states $\bar{Q}(\vec x) U_s(\vec x, \vec y) Q(\vec y)$ for a static color anti-color source pair located at spatial positions $\vec x$ and $\vec y$ respectively. The spatial Wilson line $U_s(\vec x, \vec y)=\exp(i\int_{\vec x}^{\vec y}A_\mu dx^\mu)=\prod U_\mu$ is a path-ordered product of link variables from $\vec x$ to $\vec y$. We replace the spatial part of trial states in each time-slice with an alternative operator constructed from eigenmodes $v_i(\vec x)$ of the three-dimensional gauge-covariant lattice Laplace operator $\Delta$, 
\begin{eqnarray}
\Phi(\vec x, \vec y) & = & \bar{Q}(\vec x)\sum_{i=1}^{N_v}\rho_i^2v_i(\vec x)v_i^\dagger(\vec y) Q(\vec y) \,,
\label{eq:lts}
\end{eqnarray}
which respects the gauge transformation behavior of the spatial Wilson line and ensures gauge covariance of the trial state. We denote \eq{eq:lts} as a {\it Laplace trial state}, where we include a quark profile $\rho_i$, which modulates the contribution from different eigenmodes. 
In \cite{Hollwieser:2022doy} we confirmed the consistency of results from classical Wilson loops and (temporal) {\it Laplace trial state correlators}. 
We improve the overlap of the operator by introducing a set of Gaussian profile functions $\rho(\lambda_i)=e^{-\lambda_i^2/4\sigma_k^2}$ into the the correlators and solving a generalized eigenvalue problem (GEVP) to identify the optimal trial state profiles $\tilde\rho_R^{(n)}(\lambda)$ for various energy levels $V_n(R)$ ($n=0,1,2,\ldots$). First, we prune the {\it Laplace trial state correlation matrix} $\mathcal L_{kl}$ using the four most significant singular vectors $u_i$ from a singular value decomposition (SVD)
at a specific $t_G$ via $\tilde{\mathcal L}_{mn}=u_{m}^\dagger\mathcal L_{kl}u_{n}$, which keeps a smaller set of distinct profiles and improves the stability of the GEVP. We perform the latter at the same $t_G$ for all spatial distances $R$: 
\begin{eqnarray}
\tilde{\mathcal L}(t)\nu^{(n)}(t,t_G)=\mu^{(n)}(t,t_G)\tilde{\mathcal L}(t_G)\nu^{(n)}(t,t_G).\label{eq:gevp}
\end{eqnarray}
From the eigenvalues or so-called principal correlators $\lim_{t\rightarrow\infty}\mu^{(n)}(t,t_G)=e^{-E_n(t-t_G)}$ we get the effective energies for a fixed $t_G$. From the generalized eigenvectors $\nu_k^{(n)}$ we can construct the optimal trial state profiles $\tilde\rho^{(n)}_R$ for the energy states provided by the GEVP, 
\begin{eqnarray}
\tilde\rho^{(n)}_R(\lambda_i)=\sum_k\nu_k^{(n)}\bar\rho_R^{(k)}=\sum_{k,l}\nu_k^{(n)}u_{k,l}e^{-\lambda_i^2/2\sigma_l^2}\,.\label{eq:propti}
\end{eqnarray}

Static hybrid potentials are characterized by the following quantum numbers $\Lambda^{\epsilon}_{\eta}$~\cite{Capitani:2018rox} \begin{itemize}
\item $\Lambda = 0,1,2,3,\ldots\equiv\Sigma,\Pi,\Delta,\Phi,\ldots$, the absolute value of the total angular momentum with respect to the axis of separation of the static quark-antiquark pair,
\item $\eta = +,-\equiv g,u$, the eigenvalue corresponding to the operator $\mathcal{P} \circ \mathcal{C}$, i.e.\ the combination of parity about the central point and charge conjugation,
\item $\epsilon = +,-$, the eigenvalue corresponding to the operator $\mathcal{P}_x$, which denotes the spatial reflection with respect to a plane including the axis of separation.
\end{itemize}

Note that for angular momentum $\Lambda > 0$ $\epsilon$ is not a good quantum number. 
In order to build hybrid Laplace trial states we introduce gluonic excitations via covariant derivatives of the Laplacian eigenvectors
$\nabla_{\vec k}V(\vec x)=\frac{1}{2}[U_k(\vec x)V(\vec x+\hat k)-U_k^\dagger(\vec x-\hat k)V(\vec x-\hat k)]$. We construct static hybrid potentials from correlation functions of Laplace trial states for $R=|\vec r|=|\vec y-\vec x|$ and $T=|t_1-t_0|$
\begin{eqnarray}
&&\Sigma_g^+(R,T)=\non\\
	&&\quad\sum_{\vec x,t_0,i,j}\big\langle\tr\big[U_t(\vec x;t_0,t_1)\rho(\lambda_j)v_j(\vec x,t_1)v^\dagger_j(\vec y,t_1)U_t^\dagger(\vec y;t_0,t_1)\rho(\lambda_i)v_i(\vec y,t_0)v_i^\dagger(\vec x,t_0)\big]\big\rangle,\\
 &&\Sigma_{u/g}^+(R,T)=\non\\
	&&\quad\sum_{\vec x,t_0,i,j,\vec k||\vec r}\big\langle\tr\big[U_t(\vec x;t_0,t_1)\rho(\lambda_j)\{[\nabla_{\vec k}v_j](\vec x,t_1) v^\dagger_j(\vec y,t_1)\pm v_j(\vec x,t_1)[\nabla_{\vec k}v_j]^\dagger(\vec y,t_1)\}\non\\
	&&\qquad\qquad\qquad\;
	U_t^\dagger(\vec y;t_0,t_1)\rho(\lambda_i)\{[\nabla_{\vec k}v_i](\vec y,t_0)v_i^\dagger(\vec x,t_0)\pm v_i(\vec y,t_0)[\nabla_{\vec k}v_i]^\dagger(\vec x,t_0)\}\big]\big\rangle,\label{eq:siug}\\
&&\Pi_{u/g}(R,T)=\Pi_\mp(R,T)=\non\\
	&&\quad\sum_{\vec x,t_0,i,j,\vec k\perp\vec r}\big\langle\tr\big[U_t(\vec x;t_0,t_1)\rho(\lambda_j)\{[\nabla_{\vec k}v_j](\vec x,t_1) v^\dagger_j(\vec y,t_1)\pm v_j(\vec x,t_1)[\nabla_{\vec k}v_j]^\dagger(\vec y,t_1)\}\non\\
	&&\qquad\qquad\qquad\quad
	U_t^\dagger(\vec y;t_0,t_1)\rho(\lambda_i)\{[\nabla_{\vec k}v_i](\vec y,t_0)v_i^\dagger(\vec x,t_0)\pm v_i(\vec y,t_0)[\nabla_{\vec k}v_i]^\dagger(\vec x,t_0)\}\big]\big\rangle,\label{eq:piug}
\end{eqnarray}
where we include Gaussian profiles $\rho(\lambda_i)=e^{-\lambda_i^2/2\sigma_k^2}$ to give different weights to individual eigenmodes $v_i$ according to their eigenvalues $\lambda_i$. Working with a number of Gaussian widths $\sigma_k$ we construct a correlation matrix and solve a generalized eigenvalue problem (GEVP) using the pruned matrix to identify the optimal trial state profiles $\tilde\rho_{\Lambda^\epsilon_\eta(R)}^{(n)}(\lambda)$ for the static energy levels of individual states at various distances, see also \eq{eq:gevp}.

We performed all our measurements on $48\times 24^3$ lattices with periodic boundary conditions except for anti-periodic boundary conditions for the fermions in the temporal direction. They were produced with the openQCD package \cite{Luscher:2012av} using the plaquette gauge action and two dynamical non-perturbatively $O(a)$ improved Wilson quarks with a mass equal to half of the physical charm quark mass. The bare gauge coupling is $g_0^2=6/5.3$ and the hopping parameter is $\kappa=0.13270$, the scale $r_0/a=4.2866(24)$~\cite{Sommer:1993ce}. All measurements were performed by our C+MPI based library that facilitates massively parallel QCD calculations. A total of $N_v = 200$ eigenvectors of the 3D covariant Laplacian were calculated as described in~\cite{Knechtli:2022bji}. When forming the correlations of the Laplace trial states, we apply one HYP2 smearing step with parameters $\alpha_1=1,\,\alpha_2=1$ and $\alpha_3=0.5$ to the temporal links~\cite{Hasenfratz:2001hp}. 
The error analysis in this work was done using the $\Gamma$ method \cite{Wolff:2003sm} with a recent python implementation (pyerror)~\cite{Joswig:2022qfe} with automatic differentiation~\cite{Ramos:2020scv}.

\begin{figure}
\centering
\includegraphics[width=.8\linewidth]{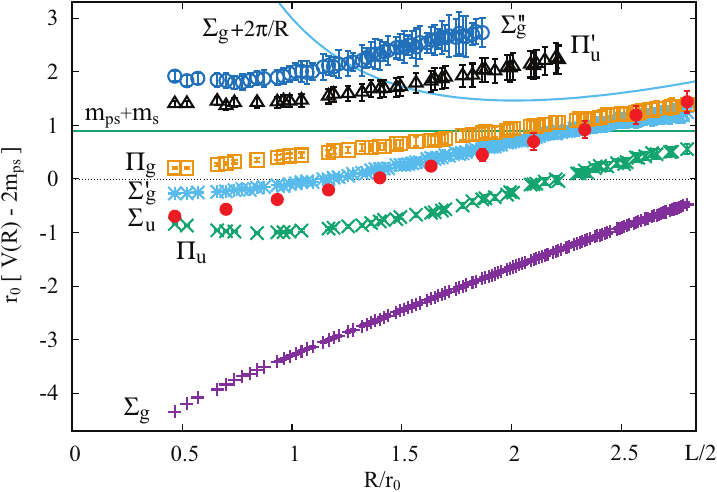}
\caption{Static hybrid potentials relative to twice the static-light S-wave mass $m_{ps}$ and the $P_-$-wave mass $m_s$ and the first radial excitation $\Sigma_g+2\pi/R$ which approaches $\Sigma_g'$. String breaking distances of $\Sigma_g^+$ and $\Pi_u$ are just above half the lattice extent. 3D resp. 2D off-axis distances for $\Sigma_g^+$ and $\Pi_{g/u}$, on-axis only for $\Sigma_u^+$.}\label{fig:states}
\end{figure}

 We show the static hybrid ground state potentials of $\Sigma_{g/u}$ and $\Pi_{g/u}$ and some excited states in \fig{fig:states}. We plot the potentials relative to twice the static-light S-wave $m_{ps}$ and also mark the $P_-$-wave mass $m_s$, also computed via Laplace trial state correlators
\bean
\hspace{-1cm}C_{S/P_-}^{sl,ij}(t)&=&\sum_{\vec x,t_0}\bigg\langle\tr_{c,d} \bigg([v_i(v_i^\dagger D^{-1}v_j) v_j^\dagger](\vec x,t_0+t;\vec x,t_0)P_\pm U_t(\vec x;t_0,t_0+t)\bigg)\bigg\rangle\\
&=&\sum_{t_0}\bigg\langle\tr_d\{\tau_{ij}(t_0+t,t_0)P_\pm\}\sum_{\vec x}v^\dagger_j(\vec x,t_0) U_t(\vec x;t_0,t_0+t)v_i(\vec x,t_0+t)\bigg\rangle
\ean
with light perambulators $\tau_{ij}=v_i^\dagger D^{-1}v_j$ from~\cite{Knechtli:2022bji} and projectors $P_\pm=(1\pm\gamma_4)/2$, see also~\cite{Bali:2010xa}.
 We plot up to half the lattice extent $L/2$ and see that string breaking distances of $\Sigma_g$ and $\Pi_u$ are just above half the spatial lattice extent. For on-axis separations the potential of $\Pi_{g/u}$ in the continuum representation can be obtained from the $E_1^\pm$ representation of $D_{4h}$. For off-axis separations we technically do not have $D_{4h}$, but we can consider off-axis separations in a 2D plane only rather than the 3d volume, to be left with one orthogonal direction for the covariant derivatives. For $\Sigma_u$ with derivatives along the separation axis we compute on-axis distances only. 
 
\begin{figure}
\centering
\includegraphics[width=.48\linewidth]{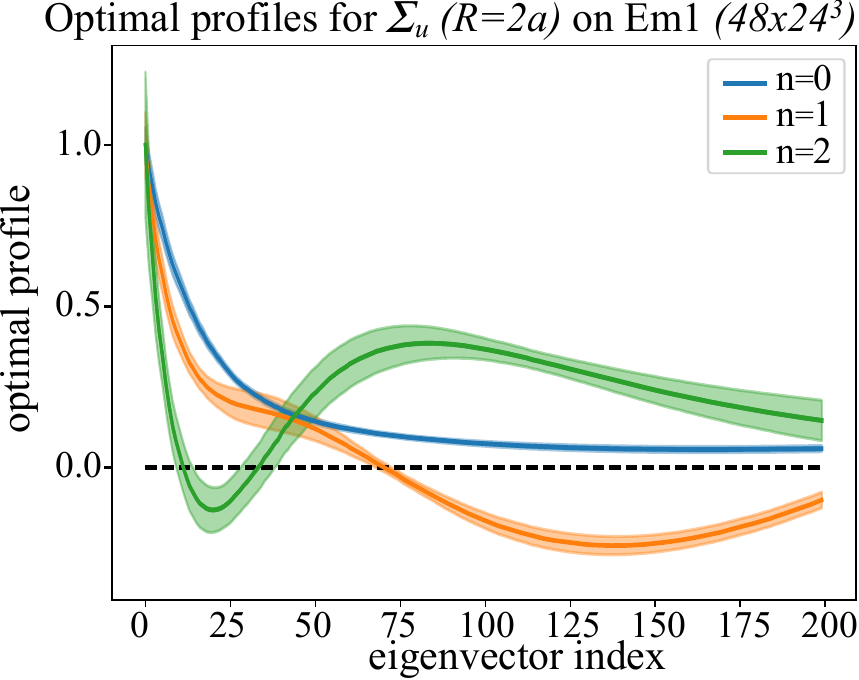}$\quad$
\includegraphics[width=.48\linewidth]{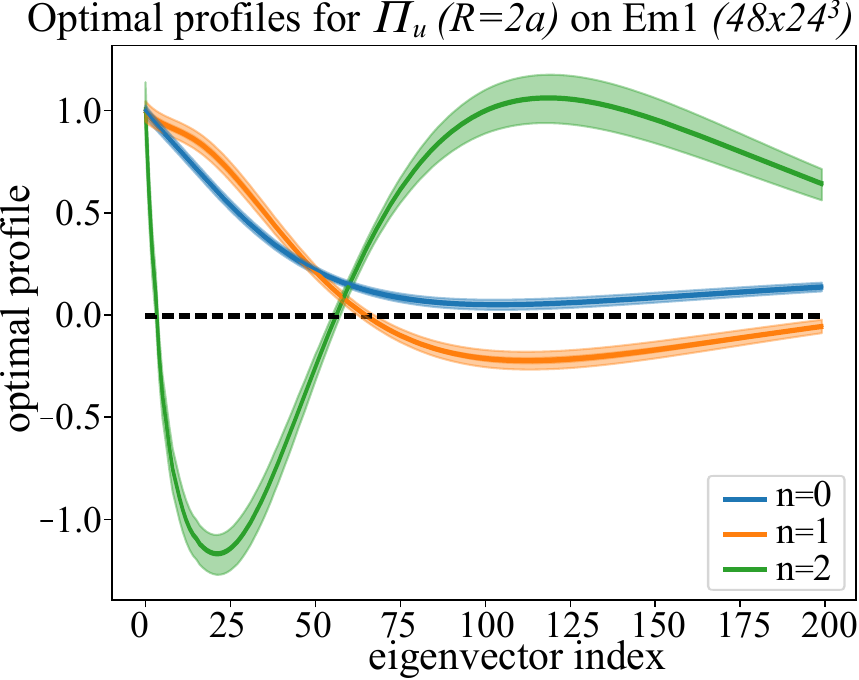}
\caption{Optimal profiles $\tilde\rho_{R=2}^{(n)}(\lambda_i)$ in \eq{eq:propti} of $\Sigma_u^+$ (left) and $\Pi_u$ (right) at distance $R=2a$.}\label{fig:propti}
\end{figure}

 In \fig{fig:propti} we show examples of optimal profiles for static hybrid energies at distance $R=2a$. The ground state profiles (blue) show that only about 100 eigenvectors are relevant. For excited states also eigenvectors corresponding to larger eigenvalues seem to play a role, but the profiles do come with somewhat larger errors and should not be over-interpreted. We can visualize a hybrid trial state by inserting an eigenvector pair $v^\dagger(\vec z)v(\vec z)$ which acts as a 'test-charge' in the Laplace trial states, {\it i.e.}, 
\begin{eqnarray}
\psi^{(n)}_{\Sigma_u}(\vec z,R)&=&\bigg\langle\sum_{\vec x,t,\vec k||\vec r}\bigg|\bigg|\sum_{i,j}^{N_v}\tilde\rho_{\Sigma_u,R}^{(n)}(\lambda_i,\lambda_j)\bigg[\nabla_{\vec k}v_i(\vec x,t)v_i^\dagger(\vec z,t)v_j(\vec z,t)v_j^\dagger(\vec x+\vec r,t)\nn
&&\qquad\qquad\qquad\qquad\quad\pm v_i(\vec x,t)v_i^\dagger(\vec z,t)v_j(\vec z,t)[\nabla_{\vec k}v_j]^\dagger(\vec x+\vec r,t)\bigg]\bigg|\bigg|_2\bigg\rangle\,,\\
\psi^{(n)}_{\Pi_{u/g}}(\vec z,R)&=&\bigg\langle\sum_{\vec x,t,\vec k\perp\vec r}\bigg|\bigg|\sum_{i,j}^{N_v}\tilde\rho_{\Pi_{u/g},R}^{(n)}(\lambda_i,\lambda_j)\bigg[\nabla_{\vec k}v_i(\vec x,t)v_i^\dagger(\vec z,t)v_j(\vec z,t)v_j^\dagger(\vec x+\vec r,t)\nn
&&\qquad\qquad\qquad\qquad\quad\pm v_i(\vec x,t)v_i^\dagger(\vec z,t)v_j(\vec z,t)[\nabla_{\vec k}v_j]^\dagger(\vec x+\vec r,t)\bigg]\bigg|\bigg|_2\bigg\rangle\,.
\end{eqnarray}
which allows to scan a 3D time-slice via the free coordinate $\vec z$\footnote{Note, if we sum over $\vec z$ due to $\sum_{\vec z}v_i^\dagger(\vec z,t)v_j(\vec z,t)=\delta_{ij}$ we recover the trial states which go into \eq{eq:siug} and \eq{eq:piug}, as well as the profile \eq{eq:propti} from \eq{eq:propt2}.}. We average/sum over the whole lattice ($\vec x,t$), which already gives a very smooth signal on a single gauge configuration. Note that we include the optimal trial state profiles from above,  
\begin{eqnarray}
\tilde\rho_{R}^{(n)}(\lambda_i,\lambda_j)=\sum_{k,l}\nu_k^{(n)}u_{k,l}e^{-\lambda_i^2/4\sigma_l^2}e^{-\lambda_j^2/4\sigma_l^2}\,,\label{eq:propt2}
\end{eqnarray}
which depend on the two eigenvalues $\lambda_i$ and $\lambda_j$. The singular vectors $u_k$ and generalized eigenvectors $\nu^{(n)}$ are derived from the SVD and GEVP respectively, for specific quark separation distances $R$, and allow us to look at the profiles for various energy states. 

\begin{figure}[h]
\centering
$\Sigma_u:\qquad n=0\qquad\qquad\qquad\qquad\qquad\quad n=1\qquad\qquad\qquad\qquad\qquad\quad n=2\qquad\qquad\;$\\
\includegraphics[width=0.32\textwidth]{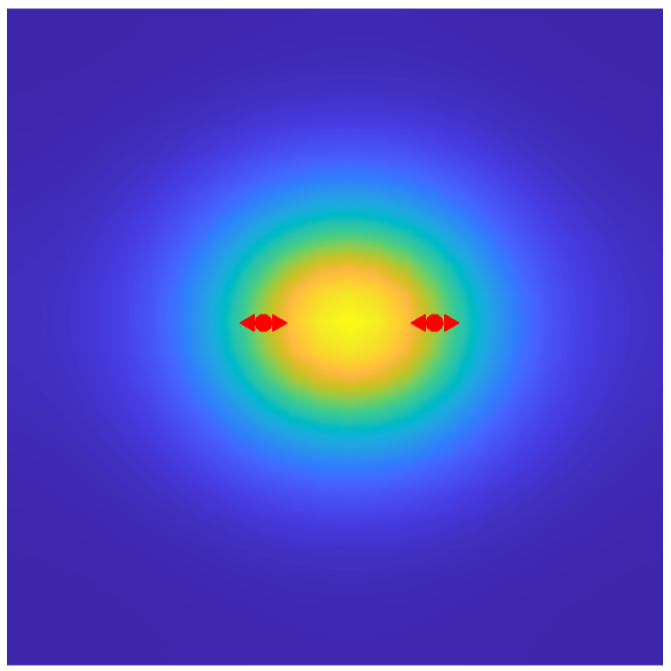}
\includegraphics[width=0.32\textwidth]{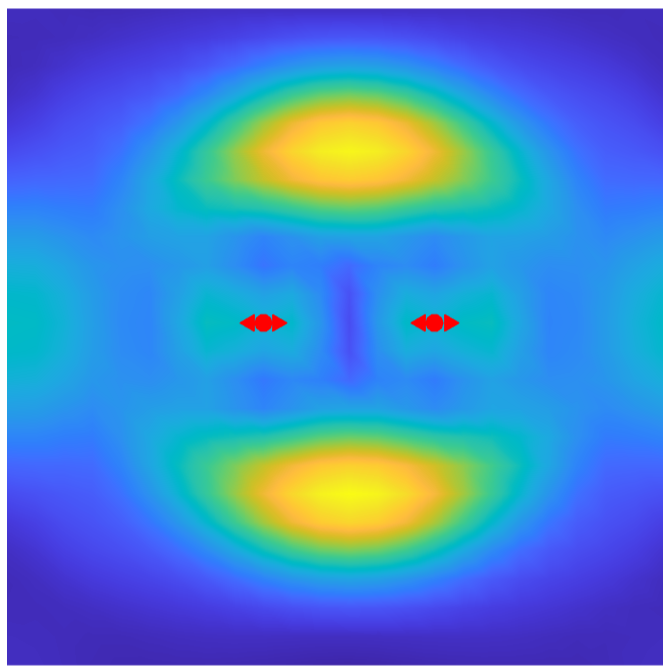}
\includegraphics[width=0.32\textwidth]{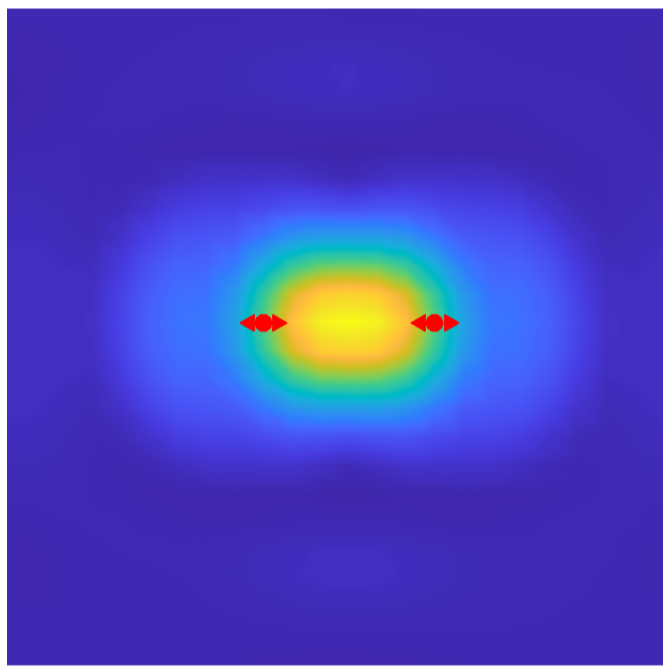}
$\Pi_u:\qquad n=0\qquad\qquad\qquad\qquad\qquad\quad n=1\qquad\qquad\qquad\qquad\qquad\quad n=2\qquad\qquad\;$\\
\includegraphics[width=0.32\textwidth]{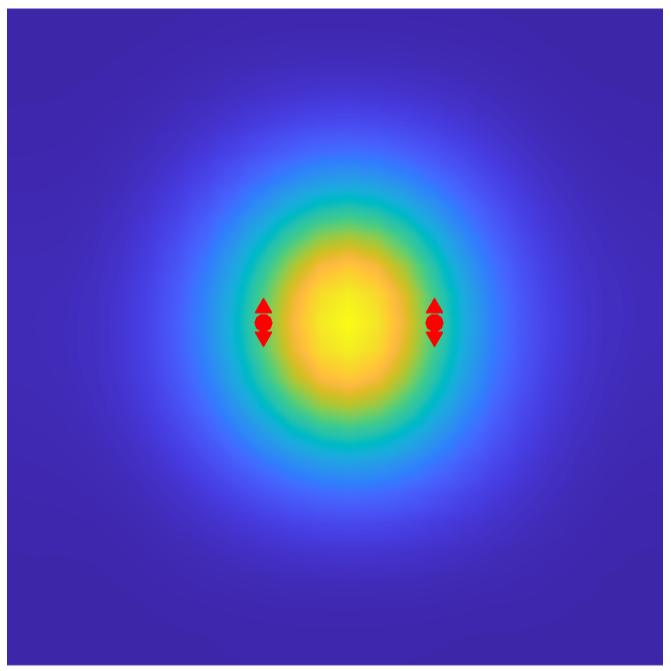}
\includegraphics[width=0.32\textwidth]{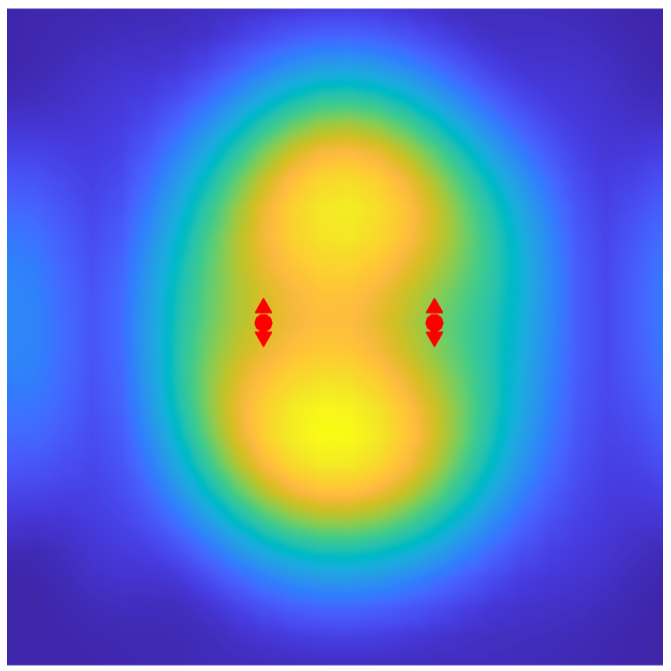}
\includegraphics[width=0.32\textwidth]{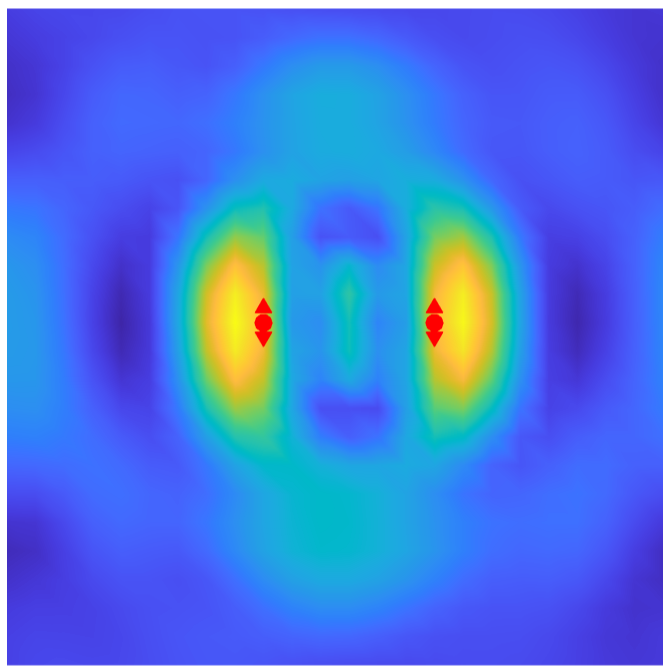}
\caption{Hybrid trial state visualizations for $\Sigma_u$  (top) and $\Pi_u$ (bottom, $\Pi_g$ almost indistinguishable) and energy levels $n=0,1,2$ (left to right) with quark-anti-quark distance $R=6a$: plane including the quark separation axis, red dots mark the static quark positions and arrows the direction of the derivatives. The color codes is yellow for the largest values of the distributions and dark blue for the smallest values.}\label{fig:prof1}
\end{figure}

In \fig{fig:prof1} we visualize the spatial distribution optimal static hybrid Laplace trial states for $\Sigma_u$ and $\Pi_u$ energy levels $n=0,1,2$ with quark-anti-quark distance $R=6a$ in a plane including the quark separation axis. Red dots mark the static quark positions and arrows the direction of the derivatives. The visualizations of excited states show additional nodes in the spatial distribution along and perpendicular to the quark separation axis, resulting from the optimal profiles. The opposite direction of the derivatives causes quite different signals for $\Sigma_u$ and $\Pi$ states, $\Pi_u$ and $\Pi_g$ however are almost indistinguishable. The physical interpretation of these distributions in terms of the chromo-electromagnetic field strength is not straightforward, the optimal profiles certainly contain some information of the ground and excited states of the static potentials, the 'test-charge' $v(\vec z)v^\dagger(\vec z)$ however does not measure a specific color field component.

\section{Conclusions \& Outlook}\label{sec:co}

We have computed static hybrid potentials $V_{\Lambda_\eta^\epsilon}(r)$ for $\Lambda_\eta^\epsilon = \Sigma_{g/u}^+$ and $\Pi_{g/u}$ states in SU(3) lattice gauge theory using alternative operators for a static quark-anti-quark pairs based on Laplacian eigenmodes, replacing traditional Wilson loops. Instead of "gluonic handles" (excitations) of the spatial Wilson lines we use symmetric, covariant derivatives of the Laplacian eigenvectors to form improved Laplace trial states by applying optimal profiles to give different weights to individual eigenvectors, derived from a generalized eigenvector problem. A high resolution of the static hybrid potentials can be achieved as off-axis distances can easily be computed in the new approach. We present a static hybrid spectrum including excited states and show their optimal profiles as well as spatial distributions of the Laplace trial states. In the spectrum we also mark the string breaking masses from static-light S- and P-waves, the string breaking distances of $\Sigma_g$ and $\Pi_u$ are just above half our lattice extent. The new methods can also be applied to multi-quark potentials. 

\section*{Acknowledgements} The authors gratefully acknowledge the Gauss Centre for Supercomputing e.V. (www.gauss-centre.eu) for funding this project by providing computing time on the GCS Supercomputer SuperMUC-NG at Leibniz Supercomputing Centre (www.lrz.de). M.P. was supported by the European Union’s Horizon 2020 research and innovation programme under grant agreement 824093 (STRONG-2020). The work is supported by the German Research Foundation (DFG) research unit FOR5269 "Future methods for studying confined gluons in QCD". The project "Constructing static quark-anti-quark creation operators from Laplacian eigenmodes" is receiving funding from the programme " Netzwerke 2021", an initiative of the Ministry of Culture and Science of the State of Northrhine Westphalia, in the NRW-FAIR network, funding code NW21-024-A (R.H.). 


\bibliographystyle{utphys} 
\bibliography{paper}

\end{document}